\documentclass{jpconf}
\usepackage{citesort}
\usepackage{amsmath}
\usepackage{amsfonts}
\usepackage{amssymb}
\usepackage{graphicx}
\usepackage{float}
\usepackage{psfrag}
\usepackage{lineno}
\usepackage{endnotes}

\newcommand{\be}{\begin{eqnarray}}
\newcommand{\ee}{\end{eqnarray}}
\newcommand{\nn}{\nonumber}

\newcommand{\yd}{{{Y^d_1}}}
\newcommand{\yydd}{{{Y^d_2}}}

\newcommand{\yu}{{{Y^u_1}}}
\newcommand{\yyuu}{{{Y^u_2}}}
\newcommand{\hggs}{\Phi_1}
\newcommand{\hggss}{\Phi_2}
\newcommand{\hgt}{\tilde{\Phi}_1}

\newcommand{\hgtt}{\tilde{\Phi}_2}
\def\dL{\mathcal{L}}
\def \Mm{\mathcal{M}}

\providecommand{\keywords}[1]{\textit{#1}}

\begin{document}

\title{Brief description of the 
flavor-changing neutral scalar interactions at two-loop level}

\author{R. Gait\'an$^1$ and J. A. Orduz-Ducuara$^2$}

\address{Departamento de F\'isica, \\
FES-Cuautitl\'an Izcalli, UNAM\\
C.P. 54740, \\
Estado de M\'exico, M\'exico.}

\ead{$^1$rgaitan@unam.mx, $^2$jaorduz@ciencias.unam.mx
}

\begin{abstract}
In this letter we show a general description about flavor-changing neutral currents (FCNC) 
mediated by scalars. The analysis is extended at two-loop level for the 
Two-Higgs Doublet Model type-III because 
others models have strong constraints on its parameters, even 
at high orders of the perturbation.
For this letter we focus on the standard model,
calculating the amplitude for 
the $h \to \gamma \gamma$ process and 
discussing the results briefly.
\end{abstract}

\keywords{To be published in Journal of Physics Conference Series (IOP). Joint Proceedings of the XV Mexican Workshop on Particles and Fields \& the XXX Annual Meeting of the Division of Particles and Fields of the Mexican Physical Society}

\section{Introduction}
The standard model (SM) of particles has awesome 
results: the experimental measurements and the theoretical 
results match very well. The observables 
are calculated using the Feynman diagrams. 
For instance, to obtain the decay width 
or the cross section, we need the Lagrangian or the  
Feynman rules (vertices and propagators  
of the model). 
In general, the SM extensions  
increase the vertex and propagators number.  
SM extensions are built to explore 
New Physics (NP) and describe unexplored phenomena. 
NP includes proposals as beyond 
standard model (BSM), 
it involves new particles, new interactions 
and new dimensions; for explaining or responding 
to the unsolved questions.

We can extend the gauge, fermion or scalar sector. 
Focusing on the last one, 
the simplest possibility to extend the SM is 
the Two-Higgs Doublet Model (THDM), which introduces a 
doublet scalar field  
plus the SM doublet. 
Theoretical motivations to enlarge the 
scalar sector could explain, e.g.,  
the CP violation and 
the flavor-changing neutral currents (FCNC) 
in the gauge sector \cite{Branco:2011iw}.

The rich phenomenology for the THDM 
has been widely studied by different 
theoretical and phenomenological groups.
Some reports show interesting results 
for the Flavor-changing (FC) 
\cite{Arroyo:2013tna,HernandezSanchez:2012eg,DiazCruz:2004pj}, 
an analysis at one-loop level for the pseudoscalar 
appears in \cite{Diaz-Cruz:2014aga},
and at two loops in the lepton sector is found in 
\cite{Harnik:2012pb}.
Several experimental reports are 
dedicated to the exotic physics  
searching for new particles coming from a variety of 
models 
\cite{CMS:2016bja,Aad:2014wza,CMS:2016pkt,ATLAS:2015atl,CMS:2016zxk,Aad:2012gm,Aad:2015osa}.

The purpose of this paper is to 
further explore the FC  
mediated by scalar bosons at loop-level.
We were inspired by the recent experimental 
results (e.g., 
refs. \cite{Khachatryan:2015kon,CMS:2016qvi}) 
and for theoretical motivations on the scalar sector.
We are aware of the strong constraints on 
loop-level and its low contributions in the  
processes and, even, in the loop-level with FC mediated 
by scalar bosons; but we shall expect a more general 
description for the neutral scalar interactions and 
their constraints \cite{Kopp:2016rzs,orduzB:2016}, which are 
so-called flavor-changing neutral scalar interactions (FCNSI), 
though they can be found as flavor-changing scalar currents (FCSC). We shall analyze the amplitudes at loop-level to have a new perspective for these processes. 
As known, GIM mechanism controls the neutral currents at tree-level. It means that processes with 
flavor-changing mediated by vector bosons are 
constrained because of the orthogonality of the 
CKM matrix \cite{Agashe:2014kda,Quang:1998yw}. 
We would like to know whether it is 
similar for flavor-changing mediated by 
scalar bosons.

Our goal is to provide an analysis on 
the equation of the amplitud at two-loop level in general and for the 
SM and THDM to explain the FCNSI
and its contribution, if any,  for these kind of processes.  
We are motivated by the next generation of 
colliders where it would be possible to explore 
the couplings in the $\gamma \gamma$-processes
\cite{Ginzburg:2013hoa}. We will use the THDM-III 
because it has flavor-changing neutral current (FCNC) transitions mediated by scalar bosons even at tree-level.  We expect that this document be 
the first of a series of papers about the 
flavor-changing and scalars at high-loop level.  
 
 We organize this paper as follow: section \ref{sec:model} describes the lagrangian for the model and its systematic implementation by the computer. In section \ref{sec:results}, we show 
 the results at two-loop level for the amplitude  and 
 in section \ref{sec:dis-con} we leave a brief discussion and conclusions.

\section{Model
\label{sec:model}}

There are two versions of THDM, labeled  
as type I and type II, with invariance under  
$Z_2$ discrete symmetry which ensures CP  
conservation in the scalar sector. In the 
first case, all quarks acquire mass through  
one doublet whereas in type II one doublet gives mass to the up-type quarks while the other doublet gives mass to the down-type quarks. In the type III 
both doublets simultaneously give masses to 
all quarks.
In the THDM-III, it is possible to have 
flavor-changing considering  parameters, 
which may induce 
FCNSI; those parameters are 
free. There is FCNSI as long as  
the diagonalization of the fermion mass matrices  does not ensure 
the diagonalization of each of the Yukawa matrices  
\cite{Branco:2011iw,Sher:2000uq}.

\subsection{Lagrangian for the THDM}
For the THDM-III, the general potential is given 
by \cite{HernandezSanchez:2012eg}:

\be
V(\Phi_1 \Phi_2) &=&
\lambda_1\big(\Phi^\dag_1 \Phi_1 -v_1^2\big)^2
+
\lambda_2\big(\Phi^\dag_2 \Phi_2 -v_2^2\big)^2
+
\lambda_3\Bigg[
\big(\Phi^\dag_1 \Phi_1 -v_1^2\big)+
\big(\Phi^\dag_2 \Phi_2 -v_2^2\big)
\Bigg]^2
\nn\\
&&
+
\lambda_4\Bigg[
\big(\Phi^\dag_1 \Phi_1\big)
\big(\Phi^\dag_2 \Phi_2\big)
-
\big(\Phi^\dag_1 \Phi_2\big)
\big(\Phi^\dag_2 \Phi_1\big)
\Bigg]
+
\lambda_5
\Bigg[
Re\big(\Phi^\dag_1 \Phi_2\big)-v_1v_2
\Bigg]^2
\nn\\
&&
+
\lambda_6
\Bigg[
Im\big(\Phi^\dag_1 \Phi_2\big)
\Bigg]^2
\nn
\ee
where $\lambda_i$ are real, $v_i$'s are the vaccuum 
expectation values and $ \Phi_i$'s are 
the Higgs doublets.

The Yukawa sector for the THDM-III is
\be
 {\dL}^{THDM-III}_{YS} &=& {\yu}\overline{Q}_{L}^0 {\hgt} u_{R}^0  +
{{\yyuu}} \overline{Q}_{L}^0 {\hgtt} u_{R}^ 0
 + {{\yd}} \overline{Q}_{L}^0 {\hggs} d_{R}^0 +
{{\yydd}} \overline{Q}_{L}^0 {\hggss} d_{R}^0 + h.c.
\nn
\ee

where $Y_i$ are the Yukawa couplings and 
$
Q_L^0 = \left(
  \begin{matrix}
u_L \\
d_L 
  \end{matrix}
\right), 
\overline{Q}_L^0 = \left( \overline{u}_L,\overline{d}_L \right),
%
{\hggs} = \left(
  \begin{matrix}
\phi_1^\pm \\
\phi_1 
  \end{matrix}
\right),
{\hggss} = \left(
  \begin{matrix}
\phi_2^\pm \\
\phi_2
  \end{matrix}
\right), 
\tilde{\Phi}_j = i \sigma_2\Phi_j^*=
\begin{pmatrix}
{\phi_j}^* \\
-\phi_i^\mp 
  \end{pmatrix}
\hspace{3mm}
\mbox{and}
\hspace{3mm}
\phi_i =
 \frac{1}{\sqrt{2}}
(v_i + \phi^0_i + i\chi_i).
$
Then we obtained,
\be
{\dL}^{THDM-III}_{YS} &=&  
{\dL}^{THDM-III}_{YS-NS}
 +
{\dL}^{THDM-III}_{YS-CS}+ h.c.\nn
\ee
where the neutral-sector is 
\be
{\dL}^{THDM-III}_{YS-NS}
 &=&
   \yu \overline{u}_L{\phi_1^0}^*u_R 
+ {\yyuu}  \overline{u}_L{\phi_2^0}^*  u_R
+ {\yd} \overline{d}_L \phi_1^0 d_R
+
{\yydd} \overline{d}_L \phi_2^0 d_R 
\ee
and the charged-sector is
\be
{\dL}^{THDM-III}_{YS-CS}
&=&
+
  \yu\overline{d}_L (-\phi_2^{-} )u_R
+ {\yyuu} \overline{d}_L (-\phi_1^{-})  u_R
+ {\yd} \overline{u}_L \phi_1^{+} d_R
+ {\yydd} \overline{u}_L \phi_2^{+} d_R.
\ee

We will use the neutral sector for this paper.
In this sector the interaction between fermions and 
scalar is given by the Yukawa couplings ($Y_f^{}$); in general, 
this is 
\be
g_{hff}^{} \sim Y_{}^{f} \sim m_f^{} 
\ee
where $g_{hff}^{}$ is the coupling, which 
represents the vertex between the Higgs 
and pair of fermions. 
In the SM the coupling is given by 
\be
g_{hff}^{SM} = Y_{SM}^{f} = \sqrt{2}\frac{m_f^{}}{v} 
\ee
where $m_f^{}$ is fermion mass and $v$ is the vaccuum expectation 
value (VEV).
For the THDM, 
after the spontaneous symmetry breaking 
the mass matrix is given by
\be\label{eq:mf-dgnlsd-frmns}
m_f^{}=\frac{1}{\sqrt{2}} 
\Big(v_1^{}Y_1^{f}  + v_2^{}Y_2^{f}\Big)
\ee 
where we have two VEV's 
$v_1^{}$ and $v_2^{}$, which are relationated by:
$\frac{v_2^{}}{v_{1}^{}} = \tan\beta.$ 
In a general form, the eq. \eqref{eq:mf-dgnlsd-frmns} 
is non-diagonal, and it can be made by:
\be
\big(U^{f}_{L}\big)^{\dagger} 
m_f^{}
\big(U^{f}_{R}\big)^{}
=
{\widetilde{m}}_f^{}
\ee

It is important to note that the  
widetilde over the quantities means 
the flavor  basis.
Besides, in general, $U_{}^{}$ is ortogonal and 
$Y^f_{}$ could be complex.
So it is possible to have mixing between the mass (fermions) 
eigenstates at tree level. 
Considering the Yukawa matrices Hermitian, 
the mass eigenstates are
\be
f = U_f^{\dagger} f^\prime
\ee
where $f$ can be $u, d-$type quarks 
and $l$ leptons. Then the mass matrix is given 
by
\be
{\widetilde{m}}_f^{}=\frac{1}{\sqrt{2}} 
\Big(v_1^{}{\widetilde{Y}}_1^{f}  + 
v_2^{}{\widetilde{Y}}_2^{f}\Big)
\ee 
where, ${\widetilde{Y}}_i^{f} = U^\dagger_L Y^{f}_{i} U_R^{}.$

In order to reduce the free parameters, one can rewrite the 
eq.\eqref{eq:mf-dgnlsd-frmns} as:
\be
{\widetilde{Y}}_1^{f} = \sqrt{2} 
\frac{m_f^{}}{v_{}^{}\cos\beta} - 
{\widetilde{Y}}_2^{f} ~\tan\beta. 
\ee 
\noindent
Sometimes Yukawa couplings are defined in terms on 
${\widetilde{\chi}}_{ij}^{}$ parameters 
\cite{Diaz-Cruz:2014aga}; namely,
\be
{\widetilde{Y}}_{ij}^{f} = 
\sqrt{2} \frac{\sqrt{m_i m_j}}{v}{\widetilde{\chi}}_{ij}^{f},
\ee
where $m_i$ and $m_j$ are the fermion masses, and 
${\widetilde{\chi}}_{ij}^{f}$ are free parameters. 
This specific pattern is known as Cheng \& 
Sher ansatz \cite{Cheng:1987rs}. 
Through this 
mechanism it is possible to have 
FCNSI at tree-level. \nocite{Gaitan:2015hga}
This relation has been used in different papers; e.g. 
\cite{Arroyo:2013tna,GomezBock:2005hc,Diaz-Cruz:2014aga}.

In the next subsection we show the scheme to obtain the amplitude.

\subsection{Methods}

\begin{figure}[H]\centering
\includegraphics[scale = 0.15]{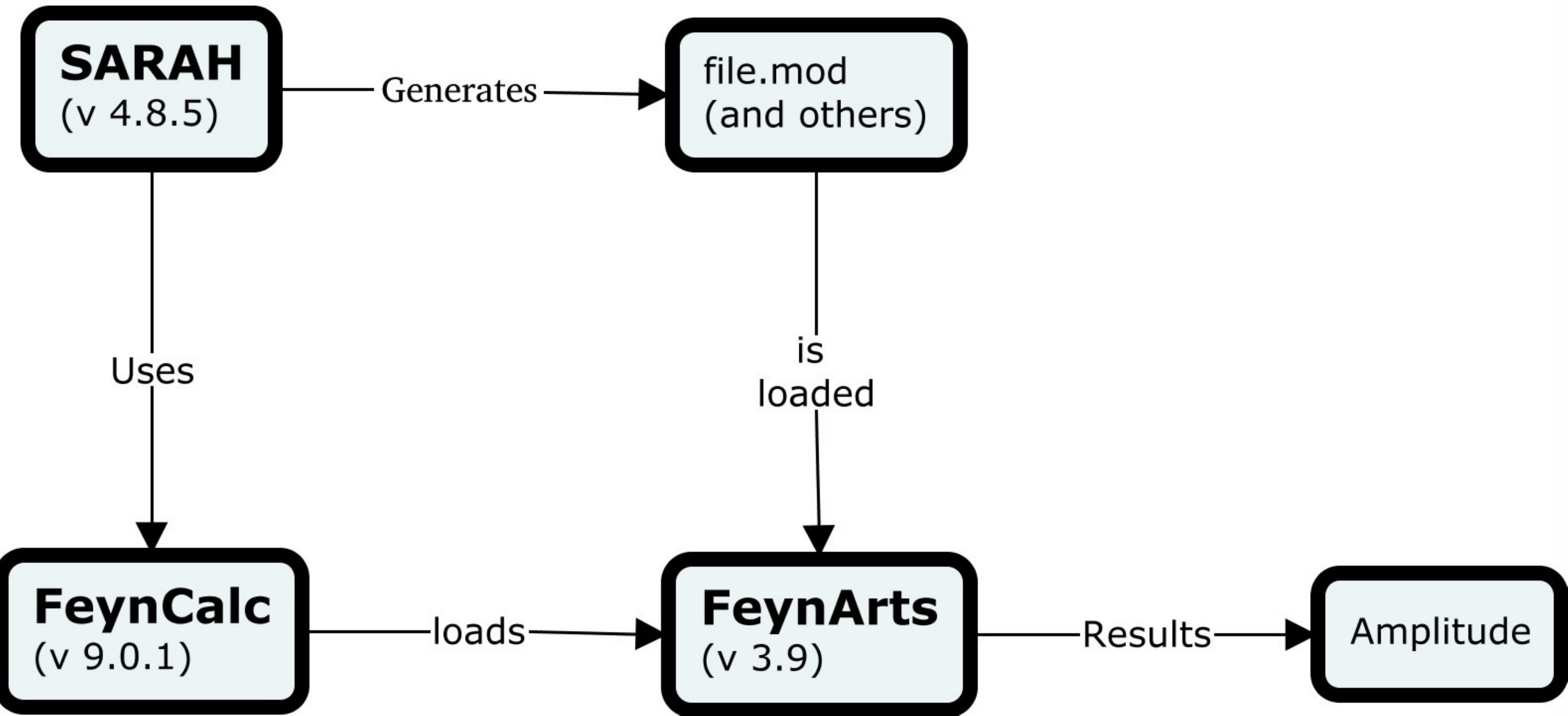}
\caption{\label{fig:Scheme-Mthds}
This scheme shows the path to follow for 
obtaining the amplitude at two-loop level, using 
SARAH
\cite{Staub:2015kfa,Staub:2013tta,Vicente:2015zba}, FeynCalc and FeynArts \cite{Hahn:2000kx}. We implemented 
the THDM-III and FV and use different 
commands to exclude field points, isolating 
a process with Higgs boson, 
which has $m_h \sim 126$ GeV, and mediated 
by virtual Higgs boson.}
\end{figure}

We calculate the amplitude at two-loop level.
We represent the scheme in fig. 
\ref{fig:Scheme-Mthds}.

Next section shows the results for the $h \to \gamma \gamma$ 
process at two-loop level, considering the flavor-changing 
mediated by a SM-like scalar boson.  

\section{Results\label{sec:results}}

\begin{figure}[H]\centering
\includegraphics[scale = 0.45]{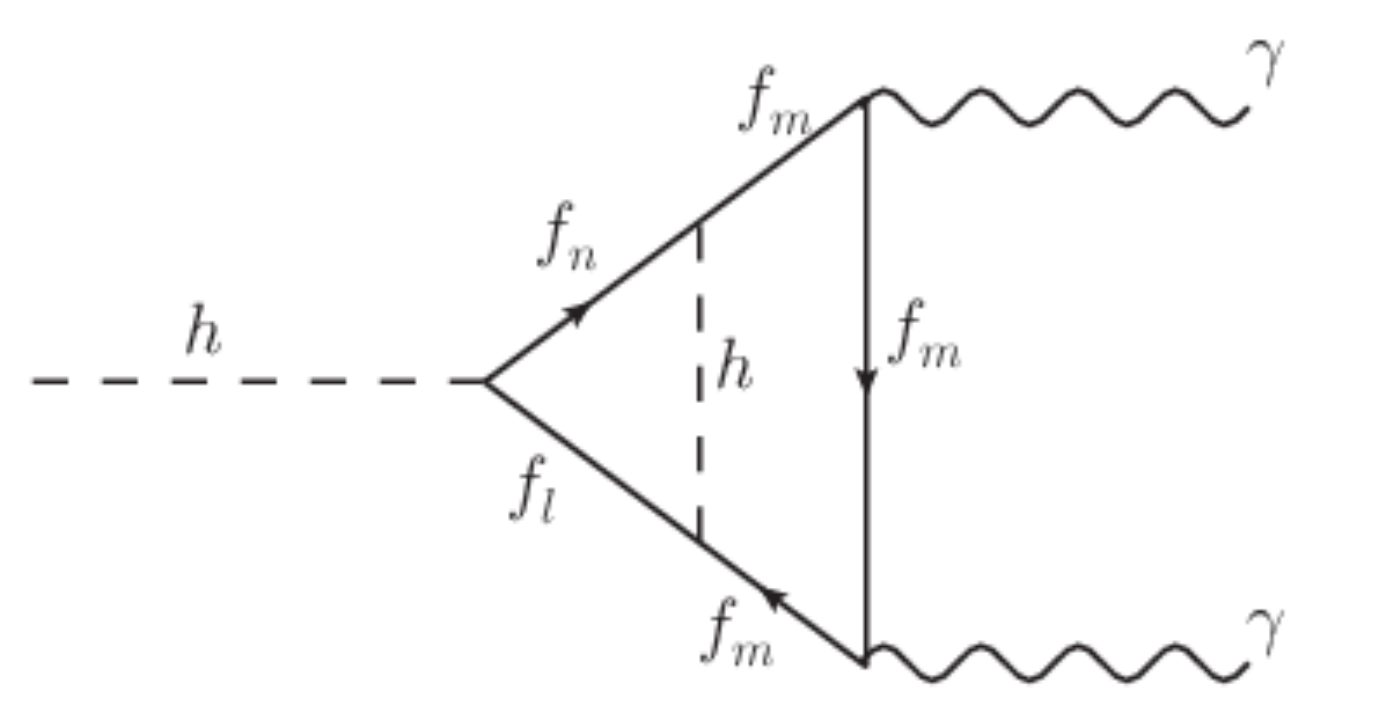}
\caption{\label{fig:all-Ampl-h-VV-two-loop} 
The Feynman diagram for the process
$h \to \gamma \gamma$ 
at two-loop level; note FC in the loop mediated 
by the scalar. We do not draw the 
adjoint process.
}
\end{figure}

We implemented the THDM-III with the 
flavor-changing in SARAH for the process represented by fig. \ref{fig:all-Ampl-h-VV-two-loop}.

The amplitude is given by
\be\label{eq:Mpltd-2L-gral}
{\Mm_{2}^{}}(\phi \to V' V) = - \frac{i}{
256 \pi^8_{}m_h^{12}\prod\limits_{\lambda=1}^{6} 
P_{\lambda}}
\mbox{Tr}{\big(M_{2}^{\mu_1^{}\mu_2^{}}}\big)
\epsilon^{*}_{\mu_1}(p_2)
\epsilon^{*}_{\mu_2}(p_3)
\ee
where the product of the propagators is
\be\label{eq:prpgtrs-2L}
\prod_{\lambda=1}^{6} 
P_{\lambda}&=&
(r_{q_1}^{}- r_{l}^{})
(r_{q_2}^{}-r_{m}^{})
(r_{p_3^{}\bar{q}_2^{}}^{}-r_{m}^{})
(r_{q_1^{}q_2^{}}-1)
(r_{p_2^{}p_3^{}q_1^{}}^{} - r_{n}^{})
(r_{p_2^{}p_3^{}\bar{q}_2^{}}^{}- r_{m}^{})
\ee
with $r_i^{}=\frac{m_i^2}{m_h^2}, 
~r_{p_i^{}\bar{p}_j^{}}^{} = \frac{(p_i^{}-p_j^{})^2}{m_h^2},
~r_{q_i^{}q_j^{}} = \frac{(q_i^{}+q_j^{})^2}{m_h^2},
~r_{p_i^{}p_j^{}q_k^{}}^{} =\frac{({p_i}+{p_j}+{q_k})^2}{m_h^2}
$
and 
$r_{p_i^{}p_j^{}\bar{q}_k^{}}^{}
=\frac{({p_i}+{p_j}-{q_k})^2}{m_h^2}.$ 
$p_1^{}$ is the momentum for the scalar boson, 
$p_{2,3}^{}$ are the momentum of the particles 
in the final state and $q_{1,2}^{}$ are the momentum for the 
loops.
The tensorial amplitud is
\be\label{eq:tnsrl-mpltd-SM-2Loop-hAA}
M_{2}^{\mu_1^{}\mu_2^{}}&=&
\kappa^{\text{SM}}
\big(m_{m}^{}+\gamma_{\nu_1^{}} 
P_{q_2^{} \bar{p}_2^{}\bar{p}_3^{}}^{\nu_1}
\big)
 \gamma^{\mu_1}(P_L^{}+P_R^{})
\nn\\
&&
\big(m_{m}^{}+\gamma_{\nu_2}^{}  
P_{q_2^{}\bar{p}_3^{}}^{\nu_2}
\big)
 \gamma^{\mu_2}(P_L^{}+P_R^{})
\nn\\
&&
\big(m_{m}^{}+\gamma_{\nu_3^{}} q_2^{\nu_3}\big)(P_L^{}+P_R^{})
\nn\\
&&
(m_{l}^{}-\gamma_{\nu_4^{}} q_1^{\nu_4^{}})({P_L^{}}+{P_R^{}})
\nn\\
&&
\big(m_{n}^{}+\gamma_{\nu_5^{}}  
P^{\nu_5^{}}_{\bar{p}_2^{}\bar{p}_3^{}\bar{q}_1^{}}
\big)
({P_L^{}}+{P_R^{}})
\ee
\noindent
where $\kappa^{\text{SM}} = (-i {e})^2\Big(-\frac{i {e} m_{f}^{} }{2 {M_W} {S_W}}\big)^3 $ and 
$P_{q_2^{} \bar{p}_2^{}\bar{p}_3^{}}^{\nu_1} = 
({q_2}-{p_2}-{p_3})^{\nu_1}_{},$
$
P_{q_2^{}\bar{p}_3^{}}^{\nu_2} = 
({q_2}-{p_3})^{\nu_2^{}},
$
and 
$P^{\nu_5^{}}_{\bar{p}_2^{}
\bar{p}_3^{}\bar{q}_1^{}} =(-{p_2}-{p_3}-{q_1})^{\nu_5^{}}_{}.$ 
In eq. \eqref{eq:tnsrl-mpltd-SM-2Loop-hAA} 
is possible to reduce the $(P_R^{}+P_L^{})$ terms, however 
these could contain the model-dependent 
parameters. It is shown simply the general 
expression for the SM, and 
the results for the THDM will be reported soon elsewhere.

\section{Discussion and conclusions 
\label{sec:dis-con}}

The discovery of the Higgs boson and exploration 
of its properties generate great interest in  
the scientific community. Theoretically, 
there are challenges to calculate 
at two- or multi-loop level associated to 
ultraviolet,  infrared and  mass 
singularities. Besides the integration for multi-loop, 
considering massive or, maybe, non-scalar integrals 
could also be awkward calculation.
Several technical calculations at one-loop 
have been developed  
\cite{Catani:2008xa,Gonsalves:1983nq,Feynman:2000fh}, 
even there are some numerical methods 
to calculate at this level  
\cite{Borowka:2015mxa,Buchta:2015wna,Cutkosky:1960sp}; 
however there are unexplored models with FCNSI at high-loop 
level. We expect that high-loop level studies can be 
interesting tests for the NP. 

In this paper, we discuss $h \to \gamma \gamma$ decay. 
This kind of process may be explored by the future 
generation of colliders. The enhanced measurements 
could show more information about the FC mediated 
by scalars and the Yukawa couplings.

{\ack 
JO thanks to postdoctoral scholarships at 
DGAPA-UNAM and RED-FAE 2015, and 
we also acknowledge support from 
CONACYT-SNI (Mexico) and the PAPIIT-IN113916 project.}

\section*{References}
\bibliography{mybilbio}

\end{document}